\documentclass[ reprint, amsmath,amssymb, aps, pra, floatfix,showkeys,showpacs]{revtex4-1}

\usepackage{graphicx}
\usepackage{dcolumn}
\usepackage{bm}
\usepackage{hyperref}

\begin{document}


\title{Comment on ``Laser cooling of $^{173}$Yb
 for isotope separation and precision hyperfine spectroscopy''}

\author{Quinton McKnight}
\author{Adam Dodson}
\author{Tucker Sprenkle}
\author{Tyler Bennett}
\author{Scott Bergeson}
 \email{scott.bergeson@byu.edu}
\affiliation{Department of Physics and Astronomy, Brigham Young University, Provo, UT 84602, USA}

\date{\today}

\begin{abstract}
We present measurements of the hyperfine splitting in the Yb-173 $6s6p~^1P_1^{\rm o} (F^{\prime}=3/2,7/2)$ states that disagree significantly with those measured previously by Das and Natarajan [Phys. Rev. A 76, 062505 (2007)]. We point out inconsistencies in their measurements and suggest that their error is due to optical pumping and improper determination of the atomic line center. Our measurements are made using an optical frequency comb. We use an optical pumping scheme to improve the signal-to-background ratio for the $F^{\prime}=3/2$ component.
\end{abstract}

\maketitle

\section{Introduction}

Measurements of isotope shifts are important for benchmarking atomic structure calculations \cite{PhysRevA.72.022503}. They can help address questions of nucleosynthesis in the early universe \cite{cowan2006}, parity non-conservation \cite{GINGES200463,Antypas2017,Flambaum2017}, measuring the charge distribution in the nucleus \cite{PhysRevLett.118.063001}, and constraining the search for new physics beyond the standard model \cite{PhysRevD.96.015011}. It is critically important, therefore, that the quality of isotope shift measurements is verified, and that systematic errors are properly identified and controlled. In the case of Yb, calculations can be extremely difficult because of significant configuration interaction in the complicated level structure \cite{0953-4075-32-5-006}.

Ten years ago, Das and Natarajan (DN) published what appeared to be definitive measurements of the hyperfine splitting in the Yb-173 $6s6p~^1P_1^{\rm o}~(F^{\prime}=3/2)$ and $(F^{\prime}=7/2)$ levels \cite{PhysRevA.76.062505}. They used laser spectroscopy on the $6s^2~^1S_0^{ } - 6s6p~^1P_1^{\rm o}$ transition at 399 nm. In a standard laser-induced fluorescence experiment using an atomic beam, these measurements are challenging because the transition in Yb-173 to the $(F^{\prime}=3/2)$ level is nearly coincident with the resonance transition in Yb-172. To overcome this problem, DN used a one-dimensional optical molasses to selectively deflect Yb-173 atoms into a spatially separated slow atomic beam. They performed standard laser-induced fluorescence spectroscopy on the now clearly-resolved Yb-173 transitions. They reported a frequency splitting of $72.093 \pm 0.036$~MHz between the $F^{\prime}=3/2$ and $F^{\prime}=7/2$ levels. This is an excellent method that appears to be plagued by spectroscopy and metrology errors. We have repeated their experiment and measure a frequency splitting at significant variance with their results.

In this comment, we will describe measurements made using two experimental systems. Both use laser-induced fluorescence on collimated atomic beams. In one experimental configuration, we use a fast atomic beam. It is generated by heating a solid Yb sample to 500 $^{\circ}$C. The beam passes through a microcapillary array \cite{senaratne2015} and is further collimated after passing through a 12 mm aperture farther downstream. This configuration is similar to an older measurement by the DN group \cite{PhysRevA.72.032506} and similar to what we used in a publication last year \cite{PhysRevA.94.052511}. In the other experimental configuration, we reproduce the optical molasses setup of DN to generate a slow (20 m/s) isotopically pure atomic beam. Our molasses laser beam has a power of 1.1 W and an intensity of 1.4 W/cm$^2$. Our probe laser beam is generated using an independent laser. We also use a fixed-frequency laser beam tuned to the $F^{\prime}=5/2$ transition to address the optical pumping problem, which we describe below.

\section{The optical pumping problem}

We question the data shown in Fig. 2 of the DN paper. That figure shows fluorescence measurements from the Yb-173 $F^{\prime}=3/2$ and $F^{\prime}=7/2$ levels in an isotopically pure, slow Yb-173 beam. Their probe laser beam size is 8 mm. The intensity of that laser beam ranges from about 0.3 to 0.5 times the saturation intensity. Rate equations show that optical pumping populates the ground state $m_F=\pm 5/2$ level after only 1~$\mu$s. Because the atoms spent 80~$\mu$s interacting with the laser beam, the $F^{\prime}=3/2$ transition should have been completely dark.

\begin{figure}
\centerline{\includegraphics[width=0.95\columnwidth]{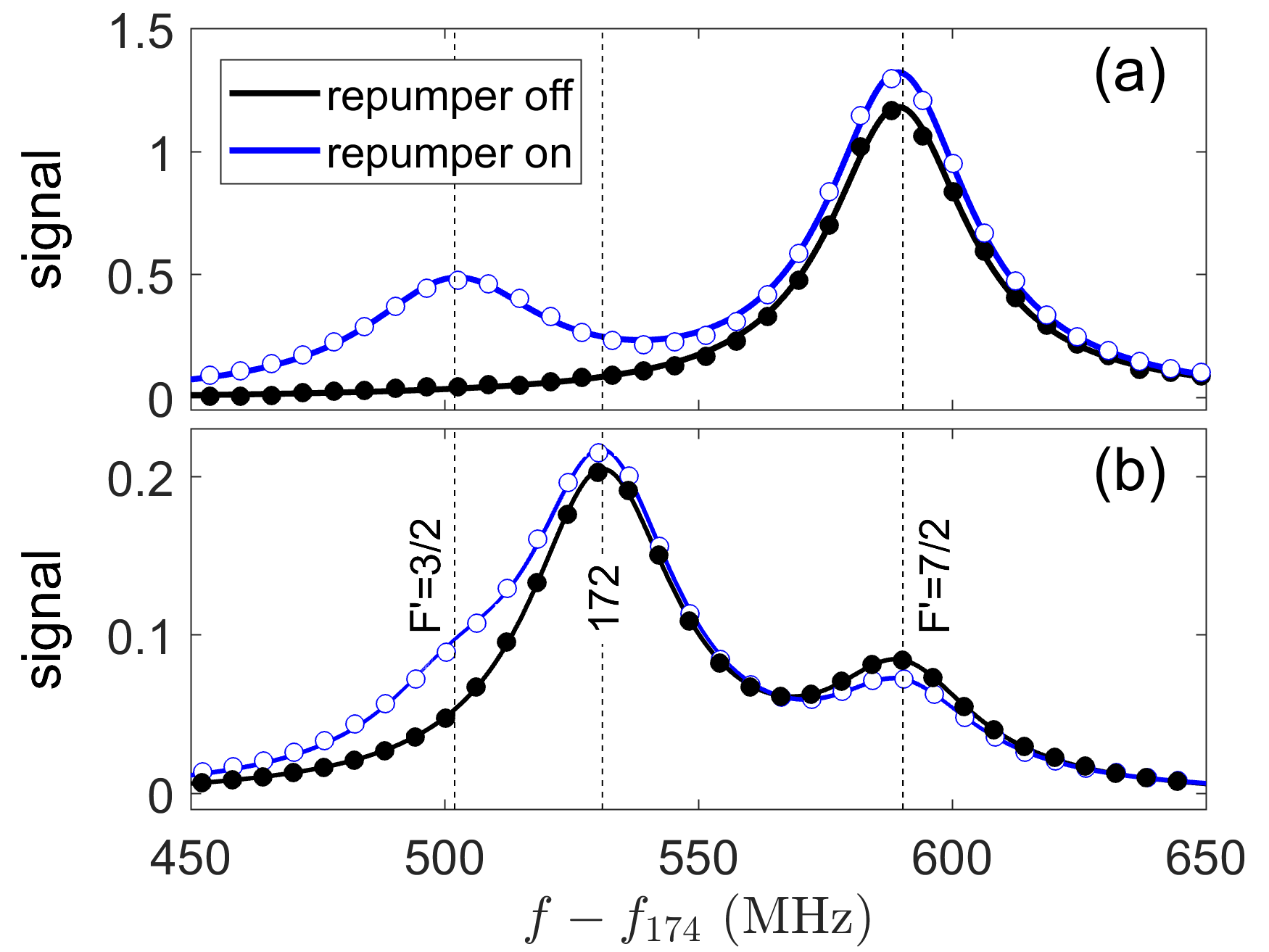}}
\caption{\label{fig:x} Fluorescence measurements from an isotopically pure, slow Yb-173 atomic beam (a) and a fast thermal beam (b). Without the repumper laser tuned to the $F^{\prime}=5/2$ transition (black circles), fluorescence from the $F^{\prime}=3/2$ level is completely absent. The repumper laser makes fluorescence from the $F^{\prime}=3/2$ level easily visible. The solid lines in this plot are fits to line shape models including 1, 2, or 3 Lorentzian peaks, as appropriate. }
\end{figure}

We show this in our optical molasses measurements. In Fig. \ref{fig:x} we show that in a nearly exact repeat of the DN measurement as they described it there is no peak from the $F^{\prime}=3/2$ level. The only peak that is visible is from the $F^{\prime}=7/2$ level.

In our experiment, We address the optical pumping problem by introducing another laser beam into the interaction region. This additional laser is tuned to the $6s^2~^1S_0^{}(F=5/2) - 6s6p~^1P_1^{\rm o}~(F^{\prime}=5/2)$ transition, approximately 840 MHz below the $F^{\prime}=7/2$ transition. This laser scrambles the population in the lower $m_F$ levels. When this laser is present, fluorescence from the $F^{\prime}=3/2$ level can be readily measured, as shown in Fig. \ref{fig:x}.

We also note that the fluorescence levels shown in Fig. 2c of the DN paper appear to be in error. That figure shows 100\% deflection of Yb-172 in the fast atomic beam. However, their optical molasses only addresses atoms with velocities less than 25 m/s. Given that the thermal velocity of their atoms is $v_{\rm th} = (k_{\rm B} T/m)^{1/2} = 165$~m/s, one would expect only $(\pi/18)^{1/2} (25/165)^3 = 0.15\%$ of the Yb-172 atoms to be deflected.

Optical pumping should have made measurement of the $F^{\prime}=3/2$ level impossible in the experiment of DN. It is not clear how the DN data could have been obtained given the information in their paper.

\section{Metrology and spectroscopy problems}

The group of DN has published many atomic transition frequency measurements over the past several years. As we pointed out in an earlier publication, their absolute transition frequencies in Yb \cite{PhysRevA.94.052511} and K \cite{falke2006} have been shown to be in error by $\sim 500$~MHz in spite of  estimated error bars of tens of kHz. Measurements in Rb \cite{maric2008} and Li \cite{sansonetti2011} deviate from frequency comb measurements in the MHz range again in spite of estimated error of tens of kHz. In all cases, issues such as quantum interference in hyperfine spectroscopy \cite{brown2013,PhysRevA.94.052511} have been neglected by them, leading to additional MHz-level errors.

Recently, the group of DN has acknowledged that their spectroscopy method has been the likely cause of errors in isotope shift and hyperfine splitting measurements \cite{1611.02406}. In many of their papers they dithered their laser by $10$ MHz while monitoring the fluorescence from their atomic samples. They demodulated the fluorescence signal at the third harmonic of the dither frequency to obtain a dispersion-shaped error signal. They locked their lasers to the zero-crossing of this error signal. Any DC-offset errors are mapped directly into a frequency error. In a recent paper, the group of DN showed that this effect was the cause of  a 4.5 MHz error in the Rb-87 $5P_{1/2}$ D1 hyperfine splitting \cite{1611.02406}.

As for the 2007 DN paper, we point out an additional error in the metrology. The spectra published in Fig. 2b of DN shows that the $F^{\prime}=3/2,7/2$ splitting is over 80 MHz, disagreeing with their final result of $72.093 \pm 0.036$~MHz. This is readily verified by extracting their data using a program such as WebPlotDigitizer and fitting to a two-Lorentzian lineshape model. DN admit that this spectrum was not the one used to determine the splitting, because they used the method described in the previous paragraph. This was necessary because their AOMs did not have enough bandwidth to tune across the line profile. Interestingly, the data in Fig. 2b of the 2007 DN paper was obtained by scanning a separate laser over the line profile, the improved method used their correction of the Rb-87 $5P_{1/2}$ D1 hyperfine splitting \cite{1611.02406}.

The laser metrology and atomic spectroscopy methods used by DN are problematic and have been shown in some cases to be in error. The Yb spectrum published by DN doesn't agree with their final result \cite{PhysRevA.76.062505}. Their final result does, surprisingly, match previous measurements from their own group \cite{banerjee2003, PhysRevA.72.032506}. It is to these measurements that we now turn our attention.

\section{Pseudo-peaks from erroneous line fitting \label{sec:bogus}}

\begin{figure}
\centerline{\includegraphics[width=0.95\columnwidth]{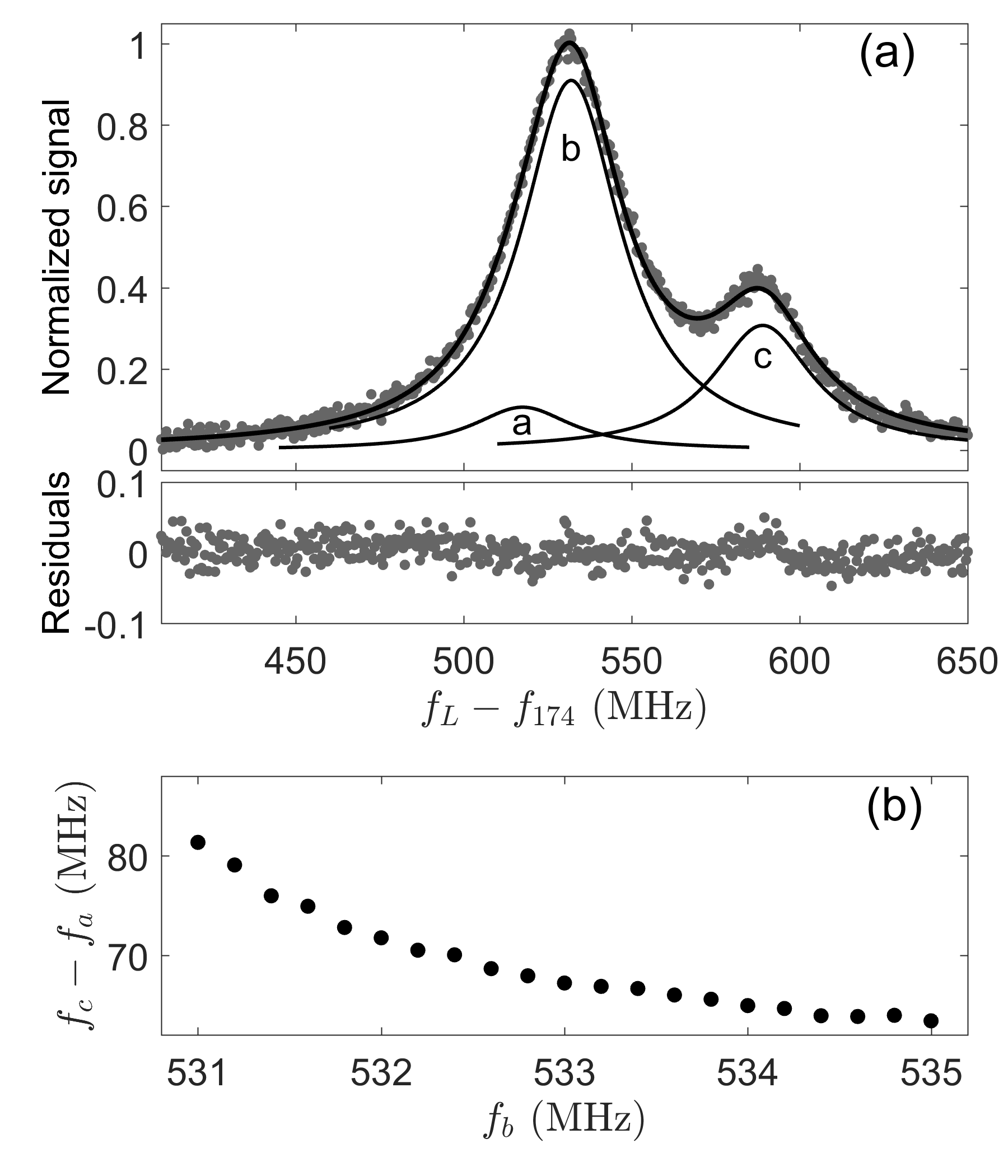}}
\caption{\label{fig:2005} SIMULATED DATA for the Yb-172/Yb-173 complex at 399 nm showing the critical error made in the analysis of Ref. \cite{PhysRevA.72.032506}. (a) The computer generates a noisy line shape consisting of two partially-resolved Lorentzians with center frequencies at 531 and 589 MHz. The dots are the simulated data, the solid line is a three-Lorentzian fit. The middle panel shows the residuals to the fit. In the model the peak at 531 MHz is somewhat broader than the peak at 589 MHz. When the center frequencies for peaks ``b'' and ``c'' are fixed, a pseudo-peak ``a'' appears in the fit. (b) The frequency difference between peak ``c'' and the pseudo-peak ``a'' as a function of the ``fixed'' frequency for peak ``b.'' Typical $1\sigma$ statistical uncertainties in determining the center frequency in repeated simulations are in the 0.3 MHz range. Dividing this by the square-root of, say, 25 measurements reduces the \textit{statistical} uncertainty in determining the line center to 0.06 MHz.}
\end{figure}

In two earlier experiments, the group of DN measured laser-induced fluorescence from Yb atoms in a fast collimated atomic beam \cite{banerjee2003, PhysRevA.72.032506}. When the laser was scanned across the Yb-172/Yb-173 complex at 399 nm, they observed two peaks. The dominant one they correctly attributed to Yb-172. The smaller one they correctly attributed to Yb-173 $(F^{\prime}=7/2)$. When they fit this signal to a two-Lorentzian line shape model, they noticed that the Yb-172 peak was 1.3 times the width of the Yb-173 $(F^{\prime}=7/2)$ peak. They hypothesized that this was due to the presence of the Yb-173 $(F^{\prime}=3/2)$ component at a somewhat lower frequency. They then analyzed their data using a line shape model with three Lorentzians, requiring the width of all three Lorentzians to be the same. Critically, they fixed the frequencies of the Yb-172 and Yb-173 $(F^{\prime}=7/2)$ transitions and allowed the frequency of the lesser component to be a fit parameter. This artificially forces a peak to appear.

We show the unsuitability of this fitting procedure using a computer simulation. We simulate their data analysis by generating a Lorentzian peak at 531 MHz and a Lorentzian peak at 589 MHz, the approximate positions of the Yb-172 and Yb-173 $(F^{\prime}=7/2)$ transitions (see Fig. \ref{fig:2005}). We add pseudo-random noise with an rms value of 1.5\%. The simulated Yb-172 peak has a Lorentzian full-width of 39 MHz. The simulated Yb-173 $(F^{\prime}=7/2)$ peak has a Lorentzian full-width of 30 MHz, similar to what was observed in Ref. \cite{PhysRevA.72.032506}. We fit this simulated data to a three-Lorentzian line shape model, fixing the frequencies of the Yb-172 and Yb-173 $(F^{\prime}=7/2)$ transitions, and requiring the widths of all three peaks to be the same.

To show the error in this fitting procedure, we vary the frequency of the Yb-172 peak in the model. This forces a third peak to appear. In Fig. \ref{fig:2005}(b) we show the frequency of the pseudo-peak as a function of the frequency of the Yb-172 peak in the model. As can be readily seen, a very good fit can be obtained with a shift that corresponds to the published data of Ref. \cite{PhysRevA.72.032506} if one carefully chooses the ``correct'' frequency for Yb-172. This simulation shows that the analysis used in the paper of Ref. \cite{PhysRevA.72.032506} (and also Refs. \cite{banerjee2003} and \cite{xinye2010}) is completely unfounded. The fact that the results of DN agree with this earlier publication in spite of the many concerns listed above is remarkable and surprising.

\section{Frequency comb measurements}

We now present our measurements of the Yb-173 $F^{\prime}=3/2$ and $F^{\prime}=7/2$ levels. We have measured fluorescence from these levels in a fast thermal atom beam with the Yb isotopes in their natural abundances and also using an isotopically pure beam of slow Yb-173 atoms. In both cases, we used a pump laser beam tuned to the $F^{\prime}=5/2$ level to overcome the optical pumping problem discussed previously. The pump and probe beams are orthogonally polarized. They are combined on a polarizing beam splitter cube before traversing the atomic beam. The pump and probe beam intensities are less than 1.5 mW/cm$^2$. 

In our measurements, these weak probe laser beams cross a collimated Yb atomic beam at a right angle \cite{PhysRevA.94.052511}. We collect scattered laser photons in a direction orthogonal to both the laser propagation direction and the atomic beam direction. We offset-lock our probe laser to an optical frequency comb \cite{Fendel:07,Lyon:14}. We have shown that the absolute frequency error in our experiment is less than 40 kHz \cite{PhysRevA.94.052511}. In the fast-beam measurements, the interaction region is enclosed in a single-layer mu-metal shield to minimize the influence of ambient magnetic fields on our measurements. In the slow beam measurements, the lasers are retroreflected. 

For convenience, we will define the fluorescence collection direction as the $\hat{z}$-axis and measure the laser polarization angle $\theta_L$ with respect to that axis, as in previous work  \cite{brown2013,PhysRevA.94.052511}. Because the pump and probe polarizations are orthogonal to each other, we will only refer to the polarization of the probe beam. A half-wave plate located before the vacuum chamber is used to rotate the polarization of both the pump and probe laser beams. This allows us to assess the influence of quantum interference on the apparent centers of the transitions \cite{brown2013}. For some measurements, we use a chopper wheel in the pump beam and phase-sensitive detection to eliminate background signal, for example, from the Yb-172 isotope.

\begin{figure}
\centerline{\includegraphics[width=0.95\columnwidth]{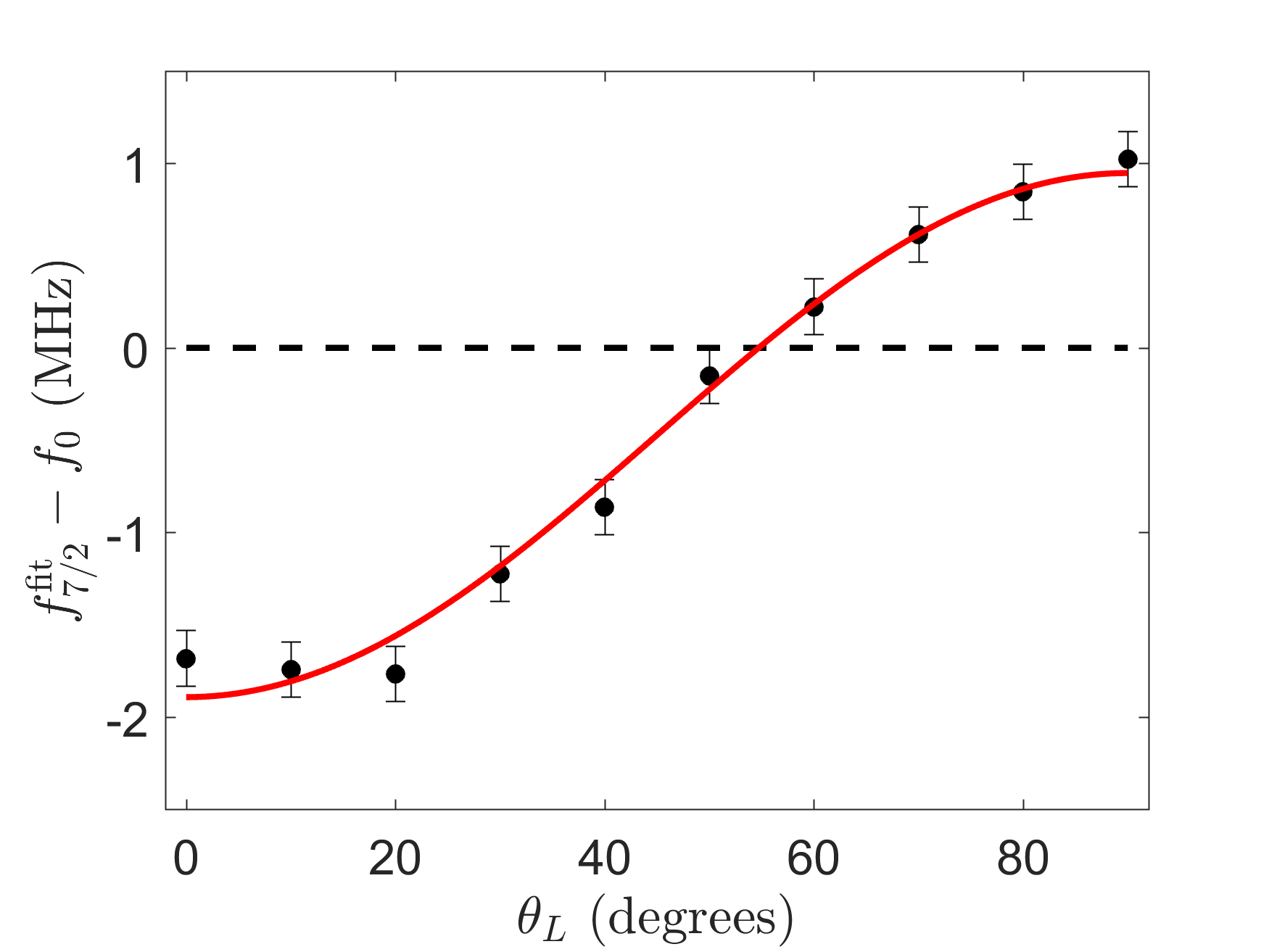}}
\caption{\label{fig:qi}
Measurements of the apparent transition frequency for the $F^{\prime}=7/2$ peak as a function of the polarization angle, $\theta_L$. Quantum interference in the excitation and decay pathways shifts the peak. Measurements made at $\theta_L = 54.7^{\circ}$ eliminate this effect.
}
\end{figure}

Quantum interference in the excitation and decay pathways can shift the frequency of line centers measured in fluorescence spectroscopy \cite{PhysRevA.94.052511,brown2013}. Our measured frequencies of the $F^{\prime}=7/2$ level in Yb-173 are shown in Fig. \ref{fig:qi}. As we rotate the laser polarization angle, we see a 3~MHz shift in the transition frequency. We see a somewhat larger shift but with opposite sign in the $F^{\prime}=3/2$ transition. The measurements account for this shift by setting $\theta_L = \cos^{-1}(3^{-1/2}) = 54.7^{\circ}$, where the quantum interference terms vanish.

Based on these frequency-comb measurements, we report the frequencies of the Yb-173 $6s^2~^1S_0^{} - 6s6p~^1P_1^{\rm o}~(F^{\prime}=3/2)$ and $(F^{\prime}=7/2)$ transitions, relative to Yb-174, to be
\begin{equation}
  \begin{array}{cc}
    6s^2~^1S_0^{} - 6s6p~^1P_1^{\rm o}~(F^{\prime}=3/2) & 503.22 \pm 0.70~\mbox{MHz}\\
    6s^2~^1S_0^{} - 6s6p~^1P_1^{\rm o}~(F^{\prime}=7/2) & 589.51 \pm 0.33~\mbox{MHz}.
  \end{array}
\end{equation}
This value of $F^{\prime}=7/2$ is consistent with our previous measurement \cite{PhysRevA.94.052511}, differing by less than 1 standard deviation. The frequency splitting between these two hyperfine levels is $86.29 \pm 0.77$~MHz.

\section{Comparison with the literature}

We calculate the hyperfine constants for Yb-173 using the Hamiltonian in Ref. \cite{RevModPhys.49.31} and also the $F^{\prime}=5/2$ data from Ref. \cite{PhysRevA.94.052511}. Our values are given in Table \ref{tab:oldMeasurements} and compared with values from the literature.

Laser spectroscopy measurements for Yb-173 are reported in Refs. \cite{hogervorst2002, banerjee2003, PhysRevA.72.032506, PhysRevA.76.062505} and \cite{xinye2010}. The group of DN published values with the smallest error estimates in Table \ref{tab:oldMeasurements} \cite{banerjee2003,PhysRevA.72.032506, PhysRevA.76.062505}, and we question these publications in this comment. The analysis in Ref. \cite{xinye2010} follows the unfortunate analysis of the DN group, which we have shown in Sec. \ref{sec:bogus} to be prone to significant systematic error. None of the values in the literature address the quantum interference in the $F^{\prime}=7/2$ level. None of them address the influence of optical pumping for the $F^{\prime}=3/2$ fluorescence level. In the atomic beam data of Refs. \cite{banerjee2003, PhysRevA.72.032506, xinye2010}, the $F^{\prime}=3/2$ component is invisible.

The laser spectroscopy measurement of Ref. \cite{hogervorst2002} used a different approach. They measured fluorescence from an atomic beam of Yb atoms with the laser polarized at $\theta_L=0$ and $\theta_L=90^{\circ}$. In the $\theta_L=0$ measurement, the fluorescence signal is dominated by fluorescence from the highly abundant even isotopes. In the $\theta_L=90^{\circ}$ measurement, fluorescence the even isotopes are strongly suppressed compared to the odd isotopes because of the dipole radiation pattern. The small residual fluorescence from the even isotopes can be subtracted out using an appropriate scaling of the $\theta_L=0$ data. From the hyperfine constants in Ref. \cite{hogervorst2002}, we calculate a hyperfine splitting between the Yb-173 $F^{\prime}=7/2$ and $3/2$ levels to be $75.3 \pm 4.0$~MHz. This $\theta_L=0$ measurement needs to be corrected for the quantum interference effect. Our measurements suggest this is approximately 5 MHz, making their measurement a little over 80 MHz with an uncertainty of 4 MHz. This interval differs from our measurement by just under two standard deviations.

\begin{table}[b]
\caption{ \label{tab:oldMeasurements}
  Reported hyperfine constant values from the literature for the Yb-173 $6s6p~^1P_1^{\rm o}$ level.
}
\begin{tabular}{cccc}
\hline \hline
Method & year & $A_{173}$ (MHz) & $B_{173}$ (MHz) \\ \hline
 This work        & 2017   & $59.52 \pm 0.20$ & $601.87 \pm 0.49$  \\
beam \cite{xinye2010} & 2010   & $57.9 \pm 0.2$ & $608.4 \pm 0.8$  \\
molasses \cite{PhysRevA.76.062505} & 2007 & $57.693 \pm 0.006$ & $609.028 \pm 0.056$ \\
beam \cite{PhysRevA.72.032506} & 2005 & $57.682 \pm 0.029$ & $609.065 \pm 0.098$ \\
beam \cite{banerjee2003} & 2003 & $57.91 \pm 0.12$ & $610.47 \pm 0.84$ \\
beam \cite{hogervorst2002} & 2002 & $57.7 \pm 0.9$ & $602.1 \pm 1.1$ \\
level crossing$^*$ \cite{Liening1985} & 1985 & $58.1 \pm 0.3$ & $588 \pm 2$  \\
level crossing \cite{BAUMANN1977433}& 1976 & $58.45 \pm 0.80$ & $589.6 \pm 13.0$  \\
level crossing \cite{PhysRev.178.18} & 1969 & $56.9 \pm 0.50$ & $575 \pm 7$  \\ \hline \hline
\end{tabular}
$^*$Additional optical pumping was used in this experiment.
\end{table}

We acknowledge support from the National Science Foundation under grant number PHY-1500376 and from the Air Force Office of Scientific Research under grant number AFOSR-FA9550-17-1-0302.

\end{document}